\documentclass[prb,final,twocolumn,showpacs,superscriptaddress,preprintnumbers,amsmath,amssymb,floatfix]{revtex4}
\usepackage[]{graphicx}
\begin{document}
\title{{\textbf{Structural and electronic properties of LaO$_{0.85}$F$_{0.15}$FeAs superconductor modified under neutron irradiation}}}
\date{\today}
\author{A.Gerashenko}
\author{S.Verkhovskii}
\author{A.Karkin}
\author{V.Voronin}
\author{ A.Kazantsev}
\author{B.Goshchitskii}
\affiliation{Institute of Metal Physics, Ural Branch of Russian Academy of Sciences, 620041 Ekaterinburg, Russia}
\author{J.Werner}
\author{G.Behr}
\affiliation{Institute for Solid State and Materials Research IFW Dresden, 01069 Dresden, Germany}
\begin{abstract}
The effect of atomic disorder induced by neutron irradiation on the crystal structure and electronic states near $E_F$ of the lightly overdoped LaO$_{0.85}$F$_{0.15}$FeAs ($T_c=21K$) was studied by X-ray diffraction and $^{75}$As NMR. The irradiation of the polycrystalline sample by "moderate" neutron fluence of $\Phi=(0;0.5;1.6)\cdot10^{19}$cm$^{-2}$ at $T=50^{\circ}$~C leads to the suppression of superconductivity. It is shown that neutron irradiation produces an anisotropic expansion of the tetragonal lattice almost due to an increase of the Fe-As distance. A partial loss of the 2D character of the FeAs layer is accompanied with a suppression  of the gap-like feature in temperature dependence of the spin susceptibility. In the most disordered state the $^{75}$As spin-lattice relaxation rate follows the Korringa law $^{75}T_1^{-1} \sim T$, the thermal behavior being typical for an isotropic motion of the conducting electrons.
\end{abstract}

\maketitle
\smallskip

Recent discovery of high-$T_c$ superconductivity in the electron-doped oxypnictides LaO$_{1-x}$F$_x$FeAs
gave new impulse in an activity of the superconductivity researchers\cite{Kamihara_JAmChSoc.130, Sadovskii_PU51}.
Similar to superconducting cuprates, the new superconducting materials have layered structure providing an anisotropy of
electronic properties, undoped LaOFeAs compound is antiferromagnetic, and superconductivity occurs upon adding either electrons or holes in the FeAs layer. Nevertheless a pseudo gap near the Fermi energy evidenced in the photoemission spectra
\cite{Sato_JPSJpn77, Ishida_arXiv:0906.2045} and $^{19}$F , $^{75}$As, $^{57}$Fe
 NMR studies \cite{Ahilan_PRB78, Nakai_arXiv0810.4569,Grafe_arXiv0811.4508} of the electron doped LaO$_{1-x}$F$_x$FeAs superconductors, but not was observed for
the hole-doped Fe-based Ba$_{1-x}$K$_x$Fe$_2$S$_2$ superconductor. More over the pseudogap behavior of spin susceptibility becomes more
pronounced with the electron doping approaching to the optimal ($x\approx0.1$) for superconductivity in LaO$_{1-x}$F$_x$FeAs
\cite{Mukuda_arXiv0904.4301}, whereas it does in the hole-underdoped region of the cuprate superconductors. It was shown that
in the optimally and overdoped compositions of  LaO$_{1-x}$F$_x$FeAs ($x>10$) both the Knight shift and spin-lattice relaxation $T_1^{-1}$ of $^{75}$As and $^{57}$Fe respectively probing the Fermi-liquid ($\textbf{q}~\approx~0$) and staggered
($\textbf{q}(\pi,\pi)$) components of spin susceptibility $\chi(\textbf{q})$  give \cite{Terasaki_JPSJpn78} no evidence of any
$\textbf{q}$-space structure in spin susceptibility $\chi(\textbf{q})$  \cite{Grafe_arXiv0811.4508}, that might be expected in the presence of antiferromagnetic spin correlations. of any determined almost by the uniform ($\textbf{q} = 0$) component of spin susceptibility \cite{Mukuda_arXiv0904.4301}. It is suggested that pseudogap behavior of spin susceptibility is more relevant to rather complex topology of the Fermi surface including both electron and hole pockets which filling can be varied differently whether electron or holes are doped into the FeAs layer. The $^{75}$As NMR study of the overdoped  LaO$_{1-x}$F$_x$FeAs ($x=0.14$) show \cite{Nakano_arXiv0909.0318} clearly an increase of the density of states at the Fermi energy with applying an  external pressure additional to the internal "chemical" one.

\begin{figure}
\includegraphics[width=1.1\hsize]{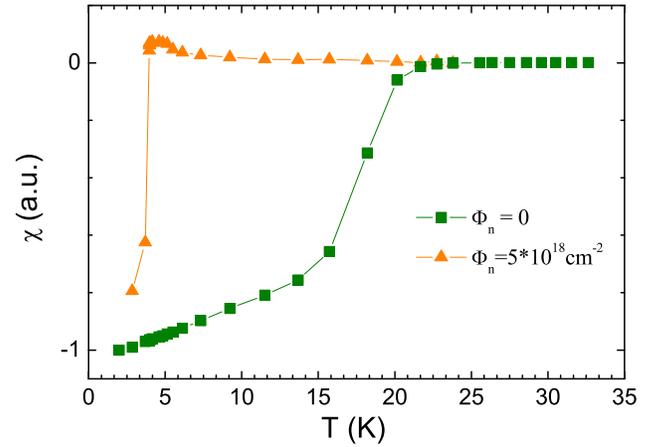}
\caption{
 The $ac$ susceptibility $\chi_{ac}$ superconducting transition curves for ordered ($\Phi=0$) and neutron-irradiated  ($\Phi=1.6\cdot10^{19}$cm$^{-2}$) samples of LaO$_{0.85}$F$_{0.15}$FeAs.
}
\end{figure}

In this report the structural X-ray and $^{75}$As NMR results are presented for light overdoped polycrystalline  LaO$_{0.85}$F$_{0.15}$FeAs ($T_c$=21K) affected by neutron irradiation. In fact the neutrons present an unique to create atomic-scale defects uniformly distributed in the lattice and acting like "negative chemical pressure" that expands crystal lattice with negligible variation of the concentration of carriers. It is shown that neutron irradiation produces an anisotropic expansion of the tetragonal lattice mainly due to the displacement of arsenic from equilibrium atomic positions, thus creating structural disorder resulting in a partial loss of the 2D character of the FeAs layer.

Polycrystalline sample of LaO$_{0.85}$F$_{0.15}$FeAs was synthesized as a pellet using two-step solid state reaction and subsequent annealing in vacuum \cite{Luetkens_PRL101}. The pellet was sliced into the plate-like samples which were irradiated with neutron fluence fluence $\Phi=(0;0.5;1.6)\cdot10^{19}$cm$^{-2}$ at $T_{irr}=50\pm10^{\circ}$~C. After irradiation both virgin and irradiated samples were moderately crushed into powder ($\sim$200mess) for X-ray, $ac$ and $dc$ susceptibility, and $^{75}$As NMR measurements. The superconducting transition temperature $T_c$ was determined as an onset of the diamagnetic response in the ac susceptibility measurements performed at driving 10~Hz magnetic field of 4~Oe with the MPMS-9 device from Quantum Design. Unirradiated sample The $ac$ susceptibility curves plotted in Fig.1 show superconducting transition of ordered sample at $T_c(\Phi=0)$=21K which shifts rather fast down to $T_c(\Phi=0.5\cdot10^{19}$cm$^{-2})$=4~K at the light structural disorder induced by neutron irradiation. The sample acquired rather moderate neutron fluence $\Phi=1.6\cdot10^{19}$cm$^{-2}$ does not show superconducting transition down to 2~K, the lowest temperature available in our experimental setup.

The structural characterization of irradiated samples was performed at room temperature by a powder X-ray diffraction technique using $Cu-K_\alpha$ radiation. The diffraction patterns obtained in the range of  $2\theta=(25-5)^\circ$ confirm a single phase tetragonal structure for each sample studied in this work. The subsequent Rietveld refinement performed within space group P4/\emph{nmm} results in structural parameters listed in Table. An insufficient broadening of the Bragg peaks observed even in the X-ray pattern of the sample with the highest neutron fluence is indicative that structural defects in the sublattices of iron, arsenic are lanthanum are almost relevant to the displacement of atoms from their positions in the ordered material.

\begin{table}[htb]
\caption
{
Room-temperature structural data of LaO$_{0.85}$F$_{0.15}$FeAs  irradiated by neutron fluence $\Phi$. Atomic posititions refined in space group $P4/nmm$ are for La$(\frac{1}{4},\frac{1}{4},z)$, Fe$(\frac{3}{4},\frac{1}{4},\frac{1}{4})$, As$(\frac{1}{4},\frac{1}{4},z)$ and O/F$(\frac{3}{4},\frac{1}{4},0)$.
}

\begin{tabular}{l|c|c|c}
\hline
\hline
Fluence   &0 &5$\times 10^{18} $cm$^{-2}$ &1.6$\times 10^{19} $cm$^{-2}$ \\
\hline
$T_c$,K       &n.s    &4  &21  \\
\hline

$a$, \AA       &4.0251(5)    &4.0251(6)  &4.0323(6)  \\
$c$, \AA       &8.7017(16)   &8.7207(16) &8.7710(18) \\
$V$, \AA$^{3}$ &140.98(4)    &141.29(4)  &142.61(4)  \\
$c/a$          &2.162        &2.167      &2.175      \\
$z_{La}$  &0.1446(7)        &0.1453(21)  &0.1457(12) \\
$z_{As}$  &0.6501(11)	    &0.6522(38)  &0.6639(20) \\
\hline
La-O/F, \AA &2.374(3)$\times$4  &2.381(9)$\times$4   &2.388(6)$\times$4  \\
Fe-Fe,  \AA &2.8462(3)$\times$4 &2.8462(3)$\times$4  &2.8513(3)$\times$4 \\
Fe-As,  \AA &2.400(3)$\times$4  &2.410(19)$\times$4  &2.477(10)$\times$4 \\
\hline
\hline
\end{tabular}
\label{tab1}
\end{table}

The main features of the radiation-induced structural disorder are shown in Fig.2. With the neutron fluence increasing a growth of the tetragonal unit cell (Fig.2b) occurs mainly along \textbf{c} direction due to the thicker Fe-As layer in irradiated samples. While the structural parameters of the La-O layer (Table I) do not show any sufficient variation. It is evident (Fig.2c) that  a thickness of the Fe-As layer is growing due to an increase of the Fe-As interatomic distance, whereas the distance \emph{d}(Fe-Fe) remains practically unchanged. Under neutron irradiation the structural changes of the nearly optimal electron-doped LaO$_{0.85}$F$_{0.15}$FeAs are developed in the way entirely opposite to those using an effect of "chemical pressure" to optimize $T_c$ in the electron-doped LaO$_{1-x}$F$_x$FeAs ($x>0.6$) \cite{Kamihara_JAmChSoc.130,Luetkens_NatMater8}  and LaO$_{1-x}$FeAs \cite{Mukuda arXiv3238,Kito_JPSJ77} compositions.
The cell volume $V$=144{\AA} of the most disordered nonsuperconducting sample is found above an upper limit of  the cell volume values which were reported somewhere  for the Fe-based superconducting pnictides \cite{Sadovskii_PU51}.

\begin{figure}
\includegraphics[width=1.1\hsize ]{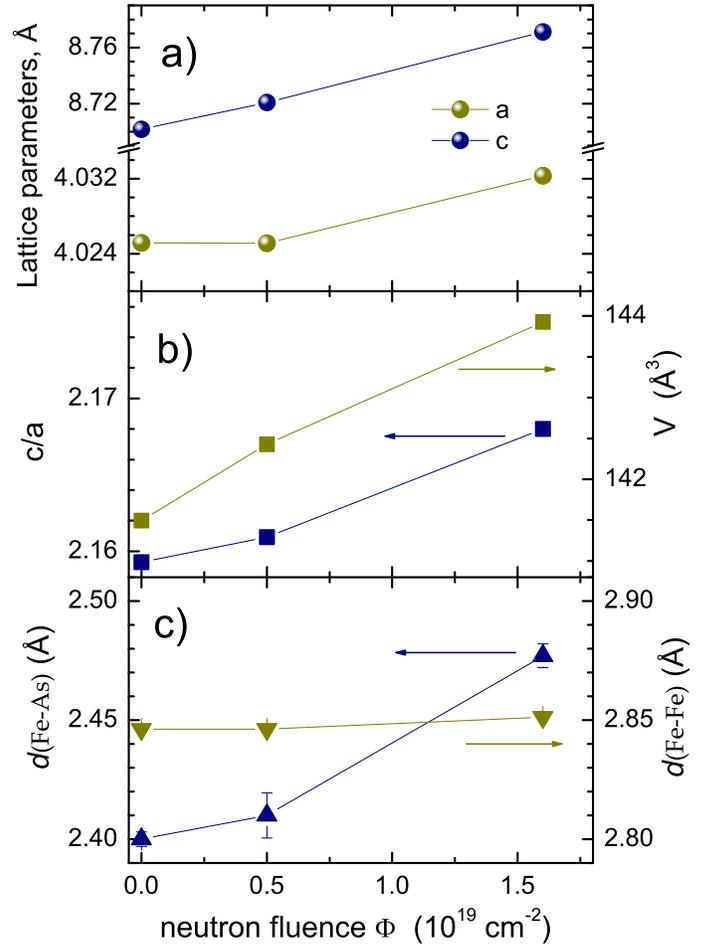}
\caption
{
 a)The tetragonal $(P4/nmm)$ lattice parameters; b) cell volume $V$, the ratio $c/a$; c) the interatomic distances $d(Fe-As)$ and $d(Fe-As)$ in the FeAs layer versus neutron fluence $\Phi$ acquired by the n-irradiated LaO$_{0.85}$F$_{0.15}$FeAs.
}
\label{fig_struct}
\end{figure}

	The $^{75}$As NMR measurements were carried out on the home-built pulse-coherent spectrometer in magnetic field of 94 kOe over the temperature range 10-300 K. Each quadrupole broadened $^{75}$As ($^{75}I$ = 3/2) NMR spectrum was obtained by summing the Fourier-transformed half-echo signals acquired at equidistant operating frequencies. Figure~3 show representative spectral patterns of the central transition ($m_I$ = -1/2 - +1/2) measured at room temperature in the LaO$_{0.85}$F$_{0.15}$FeAs  powder samples irradiated by different neutron fluence $\Phi$. The two-peaked line shape of the transition originates in an interaction of the $^{75}$As  nuclear quadrupole moment with the electric field gradient created at arsenic by electronic environment, and the high-frequency peak corresponds to the crystallites with c crystal axis oriented perpendicular ($\theta = 90^\circ$) to the magnetic field direction. It is remarkable that under neutron irradiation the peak does not show any additional broadening due apparently to induced charge disorder. This NMR observation is completely consistent with X-ray results thus evidencing that local charge symmetry at the As sites does not deviate in average from axial symmetry of the As site in the ordered ($\Phi$ = 0) material.

\begin{figure}
\includegraphics[width=1.1\hsize ]{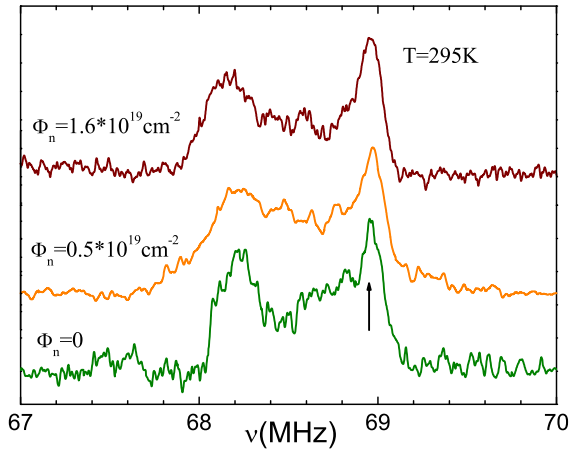}
\caption
{
Room-temperature $^{75}$As NMR spectra (transition $m =-\frac{1}{2}\leftrightarrow+\frac{1}{2}$) measured at magnetic field of 94 kOe in the powder sample LaO$_{0.85}$F$_{0.15}$FeAs irradiated by neutron fluence $\Phi$. The vertical arrow points spectral peak which at spin-lattice relaxation rate $^{75}T_1$ data were collected.
}
\label{fig_spectrum}
\end{figure}

	The $^{75}$As spin-lattice relaxation rate $T_1^{-1}$ was measured to trace thermal behavior of the spin susceptibility $\chi_s$(\textbf{q}) in the normal state of the irradiated LaO$_{0.85}$F$_{0.15}$FeAs samples. We measured the $^{75}$As spin-lattice relaxation rate $T_1^{-1}$ using an inversion recovery method. The nuclear magnetization $^{75}$m(t) was measured by integrating spectral intensity within $\pm$50 kHz around a peak ($\theta = 90^\circ$) of the central transition (Fig.~3). The recovery curve of nuclear magnetization was fitted with an expression $\{m(\infty)~-~m(t)\}~\sim~0.1{\cdot}exp(-t/T1)~+~0.9{\cdot}exp(-6t/T1$) presuming that hyperfine magnetic interaction of the nuclear spin $^{75}I$ = 3/2 with electronic spin environment is dominating.

	The temperature dependence of the $^{75}$As nuclear spin-lattice relaxation rate measured in the neutron irradiated LaO$_{0.85}$F$_{0.15}$FeAs samples is presented in Fig.~4 as a product $(T_1T)^{-1}$.  In the ordered ($\Phi$ = 0) and lightly irradiated ($\Phi=0.5\cdot10^{19}$cm$^{-2}$) superconducting samples  $(T_1T)^{-1}$ show gradual decrease from room temperature with nearly constant behavior below 30 K, which is above $T_c$. Such pseudogap behavior of $(T_1T)^{-1}$ was observed in all electron-doped Fe-based superconductors \cite{Grafe_arXiv0811.4508, Ahilan_PRB78, Nakai_arXiv0810.4569}. Following \cite{Ishida_arXiv:0906.2045} we have used an expression $a+b\cdot exp(-\Delta/T)$ to fit the $(T_1T)^{-1}$ data.
The corresponding fitting curves are plotted by solid lines in Fig.~4. As a result, the magnitude of the pseudogap is estimated to be $\Delta(\Phi=0)$=168(30)~K and $\Delta(\Phi=0.5\cdot10^{19}$cm$^{-2})$=108(20)~K with a = 0.035(6)~(sK)$^{-1}$ independent on $\Phi$ in superconducting samples. The decrease of $\Delta$ with increase of neutron fluence $\Phi$ is indicative of that  pseudogap behavior of $(T_1T)^{-1}$ originates in the specific 2D band structure near the Fermi energy \cite{ Terasaki_JPSJpn78}.

\begin{figure}
\includegraphics[width=1.1\hsize ]{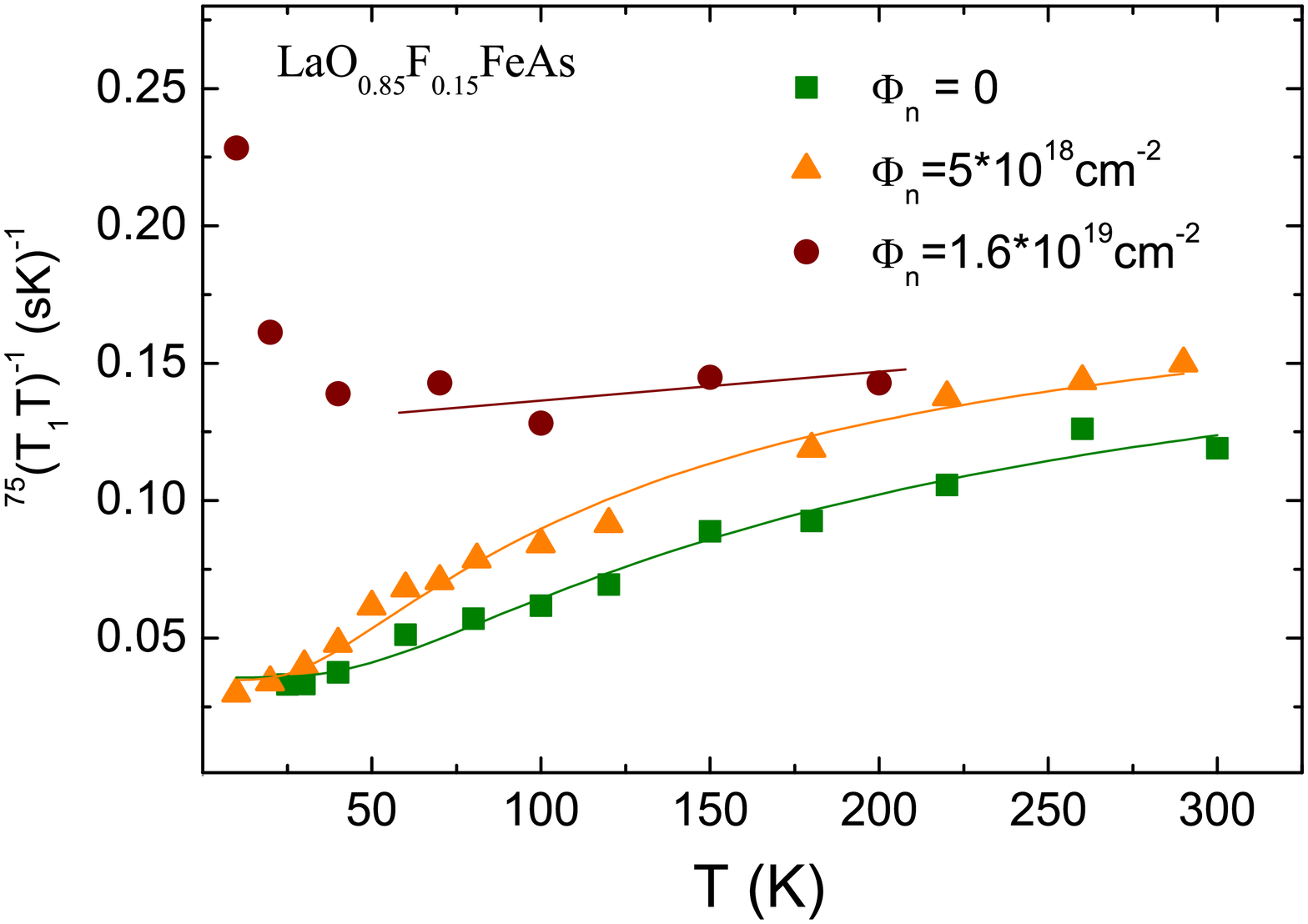}
\caption
{
The temperature dependence of the $^{75}$As $(T_1T)^{-1}$ for LaO$_{0.85}$F$_{0.15}$FeAs irradiated by neutron fluence $\Phi = 0 (\blacksquare)$, $0.5\cdot10^{19}$cm$^{-2} (\blacktriangle)$ and $1.6\cdot10^{19}$cm$^{-2} (\bullet)$. The solid curves are fits to an expression $a+b\cdot exp(-\Delta/T)$ of the corresponding $(T_1T)^{-1}$ data.
}
\label{fig_t1t}
\end{figure}

In fact, recently reported \cite{Grafe_arXiv0811.4508} scaling of $(T_1)^{-1}(T)$  measured in LaO$_{0.9}$F$_{0.1}$FeAs at $^{57}$Fe, $^{75}$As, $^{139}$La and $^{19}$F nuclei of atoms probing spin fluctuations in different areas of the \emph{\textbf{q}}-space \cite{Terasaki_JPSJpn78} gives compelling evidence, that dynamic spin susceptibility does not have any strong \emph{\textbf{q}}-dependence in the optimally electron-doped oxypnictide.

	In the most disordered nonsuperconducting sample of LaO$_{0.85}$F$_{0.15}$FeAs the $^{75}$As spin-lattice relaxation rate follows the Korringa law $^{75}T_1^{-1}\sim~T$, the thermal behavior being typical for an isotropic spectrum of the quasiparticle excitations near EF. The Curie-like upturn of $T_1^{-1}$ below 30 K is addressed to an additional contribution to $T_1^{-1}$ due accumulated structural defects, including themselves the localized magnetic moments. The magnetism of localized magnetic moments is seen clearly in the Curie term of the bulk magnetic susceptibility at low temperature. It was found that corresponding Curie constant increases proportionally to the neutron fluence.

In conclusion, an influence of structural disorder induced by neutron irradiation up to the fluence $\Phi=1.6\cdot10^{19}$cm$^{-2}$ on the spin susceptibility was studied in normal state of the lightly overdoped superconducting LaO$_{0.85}$F$_{0.15}$FeAs by measuring nuclear spin-lattice relaxation of $^{75}$As. According to the X-ray diffraction data the radiation-induced structural defects remain unchanged the tetragonal symmetry of the irradiated by neutrons LaO$_{0.85}$F$_{0.15}$FeAs. The accumulated disorder results in a growth of the cell volume, almost due to an increase of the Fe-Às interatomic distances. A partial loss of the 2D character of the FeAs layer is accompanied with a suppression  of the gap-like feature in temperature dependence of the spin susceptibility. In the most disordered state the $^{75}$As  spin-lattice relaxation rate follows the Korringa law $^{75}T_1^{-1}\thicksim T$ , the thermal behavior being typical for an isotropic motion of the conducting electrons

\begin{acknowledgments}
The authors are grateful to C. Hess, R. Klingeler, B. Buechner, A.N. Vasiliev and O.S. Volkova for valuable discussions and for providing a high quality sample of LaO$_{0.85}$F$_{0.15}$FeAs. This work is supported in part by the Programme Basic Researches RAS "Condensed Matter Quantum Physics" under project No.4 UB RAS.
\end{acknowledgments}


\begin{thebibliography}{14}

\bibitem{Kamihara_JAmChSoc.130}
Y.J. Kamihara, T. Watanabe, M. Hirano and H.Hosono, J. Am. Chem. Soc. 130 (2008) 3296.

\bibitem{Sadovskii_PU51}
M.V. Sadovskii, Physics Uspekhi 51 (2008) 122; arXiv:0812.0302.

\bibitem{Sato_JPSJpn77}
T. Sato et al., J.Phys. Soc. Jpn. 77 (2008) 073701.

\bibitem{Ishida_arXiv:0906.2045}
K. Ishida et al., arXiv:0906.2045.

\bibitem{Ahilan_PRB78}
K. Ahilan et al., Phys. Rev.B 78 (2008) 100501(R).

\bibitem{Nakai_arXiv0810.4569}
Y. Nakai et al., arXiv:0810.4569.


\bibitem{Grafe_arXiv0811.4508}
H.-J. Grafe et al., arXiv:0811.4508.

\bibitem{Mukuda_arXiv0904.4301}
H. Mukuda et al., arXiv:0904.4301.

\bibitem{Terasaki_JPSJpn78}
N. Terasaki et al., J.Phys. Soc. Jpn. 78 (2009) 013701.


\bibitem{Nakano_arXiv0909.0318}
T. Nakano et al., arXiv:0909.0318v1.

\bibitem{Luetkens_PRL101}
H. Luetkens et al., Phys.Rev. Letters 101, 097009 (2008).

\bibitem{Luetkens_NatMater8}
H. Luetkens et al., Nat. Mater. 8 (2009) 305.

\bibitem{Mukuda arXiv3238}
H. Mukuda et al., arXiv:0904.4301v2.

\bibitem{Kito_JPSJ77}
H. Kito et al., J. Phys. Soc. Jpn., 77 (2008) 063707.






\end{thebibliography}
\end{document}